\documentclass[fleqn,12pt,twoside]{article}
\usepackage[headings]{espcrc1}

\readRCS
$Id: espcrc1.tex,v 1.2 2004/02/24 11:22:11 spepping Exp $
\usepackage{graphicx}

\newcommand{\AmS}{{\protect\the\textfont2
  A\kern-.1667em\lower.5ex\hbox{M}\kern-.125emS}}

\title{Some Striking New STAR Results}

\author{C.A. Gagliardi\address{Cyclotron Institute, Texas A\&M University, College Station, TX 77843 U.S.A.}
(for the STAR Collaboration\thanks{For the full list of STAR authors and acknowledgements, see Appendix `Collaborations' of this volume.})}
       
\begin{document}

\maketitle

\begin{abstract}
Recent high-statistics Au+Au and Cu+Cu runs at RHIC have provided a wealth of new data that allow STAR to answer several outstanding questions regarding the nature of the hot, dense medium that is created in ultrarelativistic heavy-ion collisions.  However, the new data also raise new questions that require further study.  Here, we focus on a few qualitatively new results presented by STAR for the first time at Quark Matter `05.
\end{abstract}

\section{Introduction}

In its recent critical assessment of the results from the first three years at RHIC \cite{STAR_wp}, the STAR Collaboration identified several important open questions regarding the nature of the hot, dense matter that is being created in ultrarelativistic heavy-ion collisions.  During this conference, STAR has presented a broad range of important new measurements that address these questions.  These measurements are surveyed in \cite{Fuqiang,Jamie}.  Here, we focus on a few of the qualitatively new results.

At midrapidity, high transverse momentum ($p_T$) charged particle yields \cite{High_pT_incl} and back-to-back azimuthal correlations \cite{DHard03} are strongly suppressed in central Au+Au collisions, relative to those in p+p and d+Au \cite{High_pT_dAu} collisions.  These results have been taken as evidence for partonic energy loss due to induced gluon radiation \cite{jet_quench_theory} in the dense medium that is created in central Au+Au collisions at RHIC.  This suppression is so strong that it implies most of the observed high-$p_T$ hadrons originate near the surface of the collision zone and provide little information about the central density \cite{QuenchWeights}.  One needs more penetrating probes to explore the central region of the collision.  At this conference, STAR has shown first results for two such probes, heavy flavor and di-jet yields at very high $p_T$.

Conservation of energy and momentum require that the suppression of high-$p_T$ particle production must be associated with increased production at lower $p_T$.  Indeed, the back-to-back azimuthal correlations between high-$p_T$ trigger particles and low-$p_T$ associated particles are strongly enhanced and broadened in central Au+Au collisions, and the $\langle p_T \rangle$ of the associated particles approaches that of the bulk medium \cite{FWang05}.  The study of hard-soft correlations provides information both about the modification of jet fragmentation within the dense medium and the response of the dense medium to the jet.  At this conference, STAR has shown first results for three-particle, hard-soft correlations that provide detailed information about the interplay between these two effects.

\section{Heavy flavor}

\begin{figure}
  \begin{minipage}{0.49\textwidth}
    \includegraphics*[width=\textwidth]{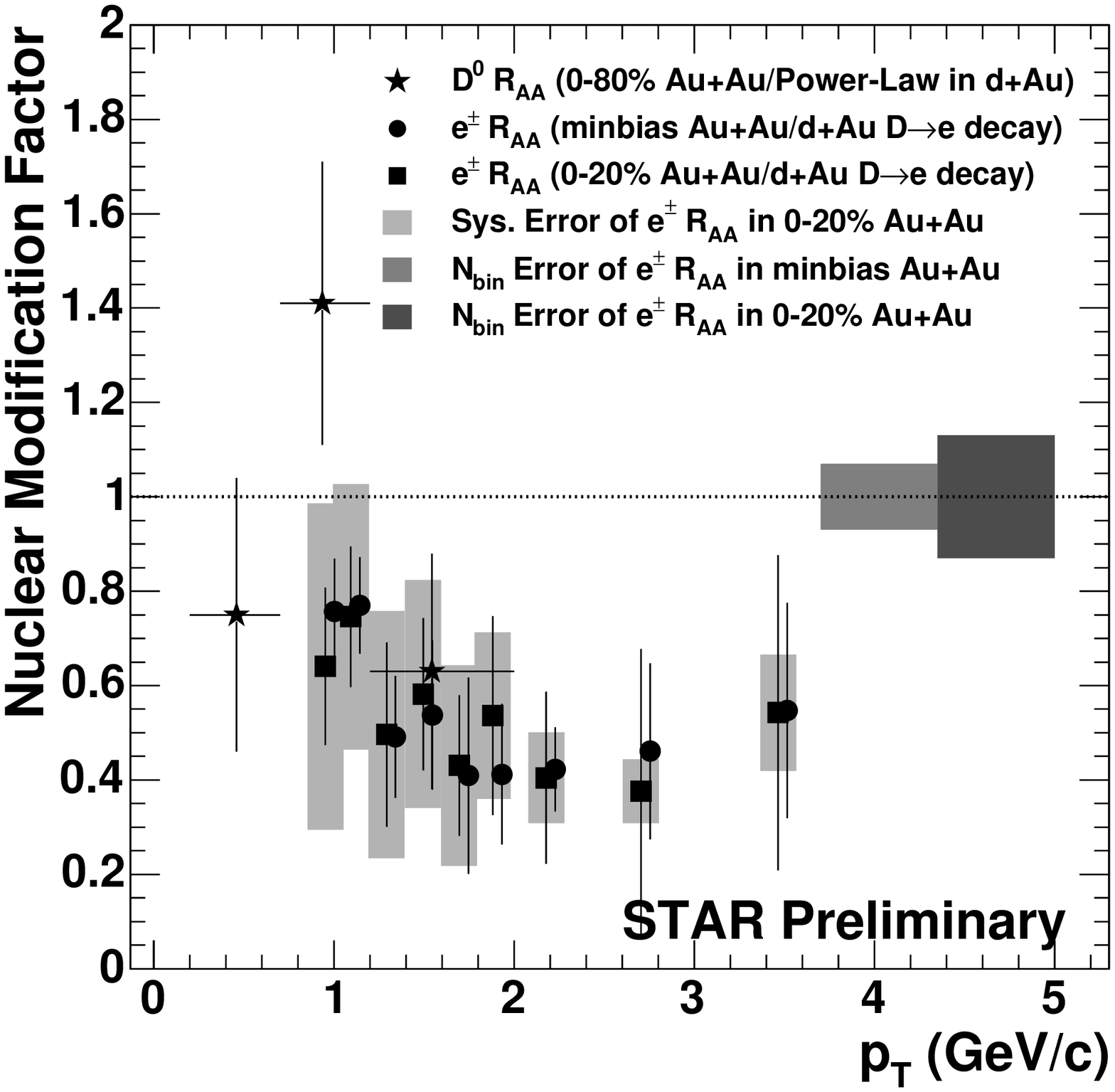}
  \end{minipage}
  \hfill
  \begin{minipage}{0.49\textwidth}
    \includegraphics*[width=\textwidth]{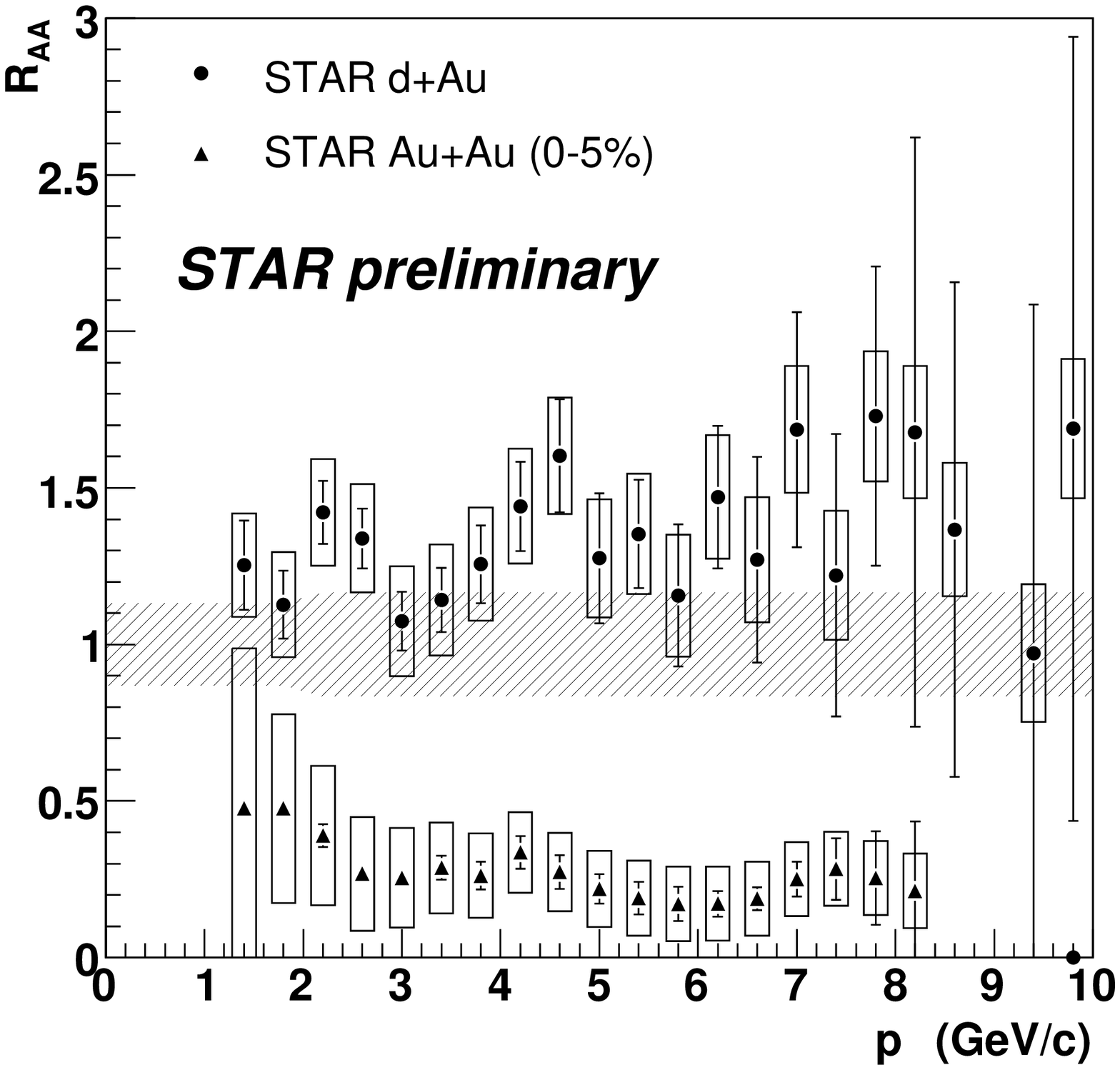}
  \end{minipage}
\caption{$R_{AA}$ as a function of $p_T$ for $D^0$ and non-photonic electrons.  The left panel shows $D^0$ and non-photonic electrons measured with the STAR TOF. The right panel shows non-photonic electrons using the STAR BEMC.  Error bars are statistical, error boxes indicate point-to-point systematic uncertainties, and the bands at unity denote the normalization uncertainties.}
\label{fig:heavy}
\end{figure}

Charm and bottom quarks, with their heavy masses, are expected to be produced early in a relativistic heavy-ion collision, and then to experience reduced coupling to the surrounding medium \cite{Doks01}.  At this conference, STAR reported the first direct reconstruction of $D^0$ mesons in Au+Au collisions \cite{Haibin} and the first measurement of non-photonic electron yields at high $p_T$ in central Au+Au collisions \cite{Jaro}.  Figure \ref{fig:heavy} shows that the yield of low-$p_T$ $D^0$ mesons in minimum-bias Au+Au collisions scales with $N_{bin}$ \cite{Haibin}, as expected.  The charm production total cross section, which is dominated by the low-$p_T$ contribution, is 1.11 $\pm$ 0.08(stat.) $\pm$ 0.42(sys.) mb/nucleon-nucleon collision, consistent with binary scaling \cite{Haibin}.  In contrast, Fig.\@ \ref{fig:heavy} also shows that the yield of non-photonic electrons, which are believed to arise from the decay of charm and bottom hadrons, is strongly suppressed in central Au+Au collisions at intermediate \cite{Haibin} and high $p_T$ \cite{Jaro}.  The suppression, a factor of four to five relative to pp collisions for $p_T > 5$ GeV/c, is comparable to that observed for inclusive charged hadrons at $p_T > 6$ GeV/c \cite{High_pT_incl}.  This contradicts theoretical expectations that heavy flavor hadrons should be less suppressed than light hadrons due to the `dead cone' effect \cite{Doks01}.  Previous energy loss calculations based on gluon radiation \cite{Arme05,Djor05} predicted that $R_{AA}$ for non-photonic electrons would saturate at 0.5 $-$ 0.6.  The calculations are only able to reproduce the observed suppression if they assume extreme medium densities -- $\hat{q}$ = 14 GeV$^2$/fm \cite{Arme05} or $dN_g/dy$ = 3500 \cite{Djor05} -- and negligible contributions from bottom for $p_T <\sim 8$ GeV/c.  These results indicate that collisional processes that have previously been neglected relative to induced gluon radiation may make an important contribution to partonic energy loss, especially for heavy quarks \cite{Coll}.

\section{Di-jet correlations at high $p_T$}

\begin{figure}
  \begin{minipage}{0.65\textwidth}
    \includegraphics*[width=\textwidth]{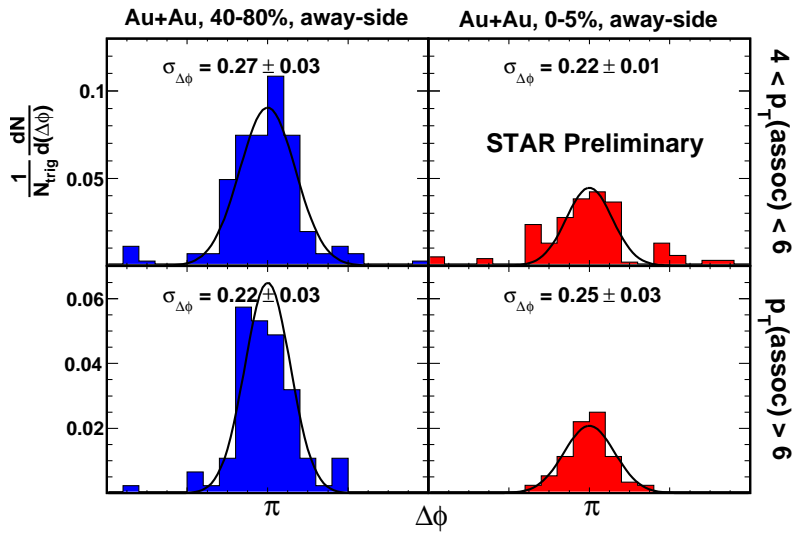}
  \end{minipage}
  \hfill
  \begin{minipage}{0.31\textwidth}
\caption{Azimuthal distributions of away-side charged hadrons for $8<p_T^{trig}<15$ GeV/c in two different centralities of 200 GeV Au+Au collisions.  These are the total observed distributions; no backgrounds have been subtracted.}
\label{fig:dphi_dists}
  \end{minipage}
\end{figure}

\begin{figure}
  \begin{minipage}{0.4\textwidth}
    \begin{center}
      \includegraphics*[width=0.88\textwidth]{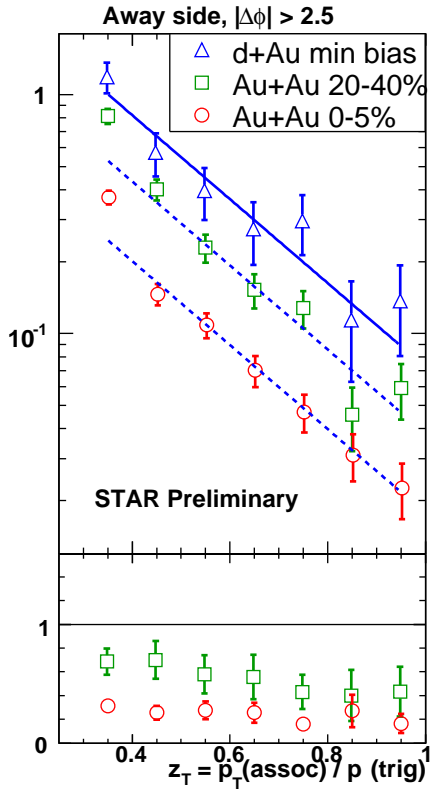}
    \end{center}
  \end{minipage}
  \hfill
  \begin{minipage}{0.56\textwidth}
\caption{(color online) Top:  Hadron-triggered fragmentation functions, $dN/dz_T$, for away-side charged hadrons as a function of $z_T$ for $8<p_T^{trig}<15$ GeV/c in 200 GeV d+Au and Au+Au collisions.  The solid line is an exponential fit to the $z_T$ distribution for d+Au.  The dashed lines represent the same exponential fit scaled down by factors of 0.54 and 0.25 to approximate the yields in 20-40\% and 0-5\% Au+Au collisions.  Bottom:  Ratio of the hadron-triggered fragmentation functions for Au+Au/d+Au.}
\label{fig:zT_away_dists}
  \end{minipage}
\end{figure}

The high statistics of the 200 GeV Au+Au Run 4 data have permitted STAR to push the trigger and associated particle thresholds to much higher $p_T$ than has been practical in previous Au+Au di-jet studies \cite{DHard03}.  Figure \ref{fig:dphi_dists} shows that a well-defined back-to-back peak, characteristic of di-jets, is observed with negligible background in both peripheral and central Au+Au collisions \cite{Dan}.  The widths of the back-to-back peaks appear to be independent of centrality, but the coincident yield is substantially less in central Au+Au collisions than in peripheral collisions.  The latter is quantified in Fig.\@ \ref{fig:zT_away_dists}, which shows the away-side hadron-triggered fragmentation functions \cite{Wang04} for d+Au and Au+Au collisions \cite{Dan}, as a function of $z_T = p_T^{assoc}/p_T^{trig}$.  The shape of the away-side fragmentation functions is unchanged from d+Au to central Au+Au, as shown by the flat ratios, but the yields are reduced by a factor of $\sim$\,4 in central Au+Au collisions.  The unmodified shape within the observed $z_T$ range is consistent with previous predictions, but the magnitude is smaller than expected \cite{Wang04}.  A different calculation predicted that significant energy loss, as evidenced by the Au+Au/d+Au ratio of $\sim$\,0.25, should be associated with significant broadening of the away-side hadron azimuthal distribution \cite{Vitev05}, in contrast to the data.  The simultaneous observations of significant suppression of the very high-$p_T$ back-to-back yields in conjunction with little or no modification of the away-side width or hadron-triggered fragmentation function will provide a significant challenge to partonic energy loss models.

\section{Three-particle correlations at intermediate $p_T$}

\begin{figure}[t]
  \begin{minipage}{0.5\textwidth}
    \begin{center}
      \includegraphics*[width=0.7\textwidth]{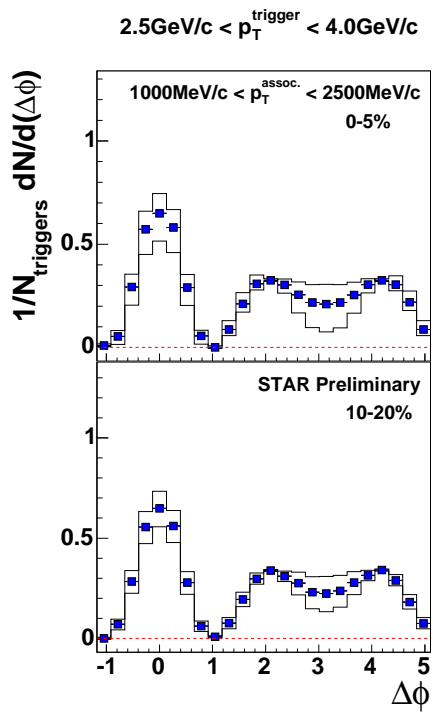}
    \end{center}
  \end{minipage}
  \hfill
  \begin{minipage}{0.46\textwidth}
\caption{Azimuthal distributions of associated charged hadrons for two different centralities of 200 GeV Au+Au collisions.  The histograms indicate the systematic uncertainty bands.}
\label{fig:int_pT_dphi}
  \end{minipage}
\end{figure}

When low-$p_T$ associated particles are observed in central Au+Au collisions opposite a higher-$p_T$ trigger particle, they are found to be significantly enhanced in number and broadened in $\Delta\phi$ relative to pp collisions \cite{FWang05}.  The $\langle p_T \rangle$ of the associated particles appears to be a minimum at $\Delta\phi$ = $\pi$ in central Au+Au collisions \cite{Jason}, in contrast to the maximum found in pp and d+Au collisions.  In addition, the away-side associated particle yield is flat or may have a small dip at $\Delta\phi$ = $\pi$ \cite{Jason}, as shown in Fig.\@ \ref{fig:int_pT_dphi}.  These novel phenomena have led to predictions that we may be observing jets that have been deflected by radial flow or a Mach cone effect associated with conical shock waves \cite{Conical}.

\begin{figure}[t]
  \begin{center}
    \includegraphics*[width=\textwidth]{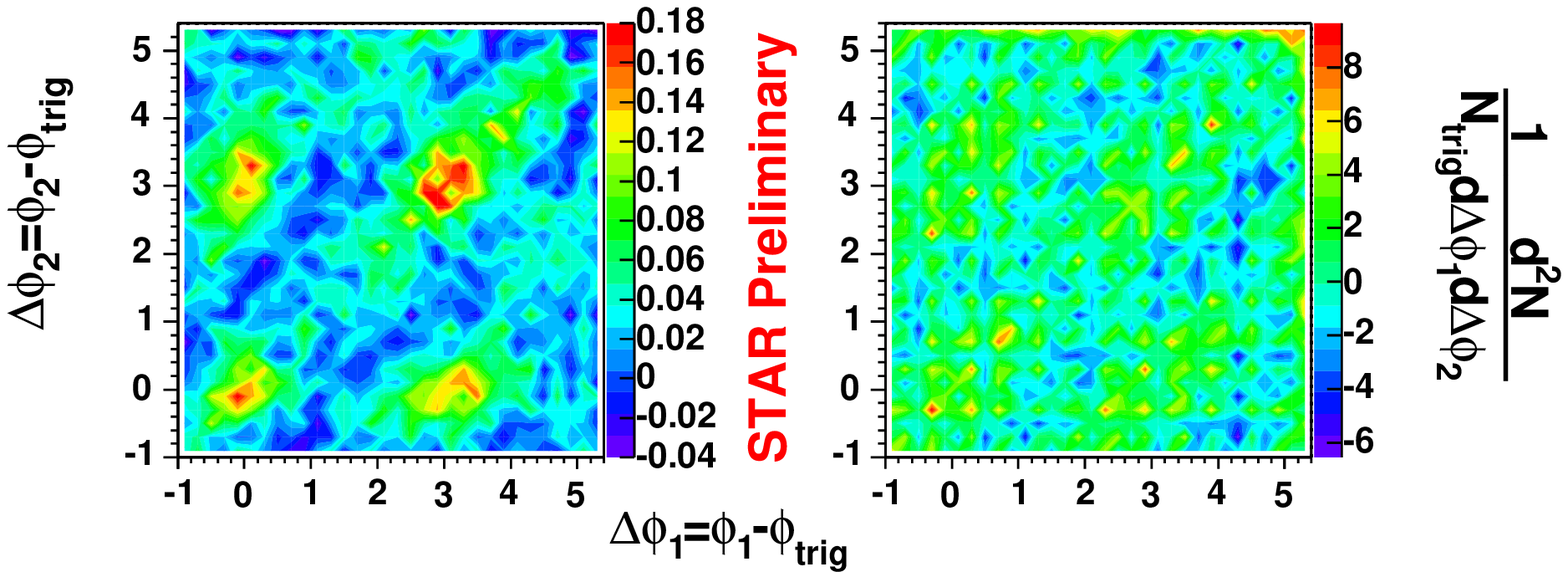}
  \end{center}
\caption{(color online) Three-particle correlations in minimum bias d+Au and 10\% central Au+Au collisions at 200 GeV.  The $p_T$ ranges are $3<p_T^{trig}<4$ GeV/c for the trigger particle and $1<p_T^{assoc}<2$ GeV/c for the two associated particles.}
\label{fig:three_part}
\end{figure}

STAR has explored three-particle correlations \cite{Jason} in the intermediate-$p_T$ region where these novel phenomena appear to be most pronounced in order to identify the underlying physical processes.  Figure \ref{fig:three_part} shows the result for minimum bias d+Au collisions and central Au+Au collisions when the true three-particle component is isolated by subtracting off the backgrounds due to all combinations of two associated particles in random coincidence with a third, in addition to the contribution from triple random coincidences \cite{Fuqiang}.  Both distributions contain the four peaks expected from di-jet production, and the peaks near $\Delta\phi_1 = \Delta\phi_2 = \pi$ appear broadened along the diagonal.  In d+Au collisions, this broadening may arise from $k_T$ effects that produce acoplanarity between partner jets.  In Au+Au collisions, an enhanced yield along the diagonal may arise from $k_T$ effects or deflected jets.  Conical flow should enhance the off-diagonal yield, for example near $(\pi \pm 1,\pi \mp 1)$.  However, for these $p_T$ ranges we find the off-diagonal coincidence yield to be substantially less than that along the diagonal \cite{Jason}, in contrast to this expectation.

\section{Conclusion}

During this conference, STAR has shown that the yields of high-$p_T$ non-photonic electrons, which are believed to arise from heavy flavor, are strongly suppressed in central Au+Au collisions.  The suppression is comparable to that observed for light hadrons.  Back-to-back di-jets have been observed unambiguously in central Au+Au collisions at very high $p_T$.  The yield relative to d+Au is suppressed by a factor of $\sim$\,4, similar to the suppression of inclusive particle yields, but the fragmentation function and the width of the back-to-back azimuthal distribution appear unmodified within the measured region.  Together, the heavy flavor and di-jet measurements will provide a severe test of our understanding of partonic energy loss in hot, dense matter, and may lead to an upper bound on the energy density of the matter as needed to obtain a lower limit on the number of degrees of freedom \cite{Mull05}.  Three-particle correlations at intermediate $p_T$ have been measured in central Au+Au collisions to explore the response of the medium when probed by an energetic parton.  The present data show the pattern expected from $k_T$ effects or deflected jets, while the characteristic signature of conical flow is not observed.

STAR has presented a wealth of new data at this conference that provide answers to several of the questions that were raised in the recent STAR critical assessment of the RHIC results from Runs 1, 2, and 3 \cite{STAR_wp}.  However, the new data also raise new questions.  For example, it is not clear whether the new results are consistent with the conventional picture that presumes partonic energy loss is dominated by induced gluon radiation.  Ultimately, the answers to all of these questions will tell us the nature of the strongly interacting matter that we are creating at RHIC.

\end{document}